\documentclass[prd, twocolumn, showpacs, showkeys, floatfix]{revtex4}

\usepackage{latexsym} 
\usepackage{amssymb}  
\usepackage{amsfonts} 
\usepackage{graphicx}       
\usepackage[dvips]{color}

\newcommand{\al}{\alpha}
\newcommand{\be}{\beta}
\newcommand{\ga}{\gamma}
\newcommand{\Ga}{\Gamma}
\newcommand{\de}{\delta}
\newcommand{\De}{\Delta}

\newcommand{\eps}{\epsilon}

\newcommand{\la}{\lambda}
\newcommand{\La}{\Lambda}

\newcommand{\si}{\sigma}

\newcommand{\Om}{\Omega}

\newcommand{\dsp}{\displaystyle}

\newcommand\eqn[1]{(\ref{#1})}      

\newcommand{\beq}{\begin{equation}}
\newcommand{\eeq}{\end{equation}}
\newcommand{\ba}{\begin{array}}
\newcommand{\bea}{\begin{eqnarray}}
\newcommand{\ea}{\end{array}}
\newcommand{\eea}{\end{eqnarray}}
\newcommand{\bc}{\begin{center}}
\newcommand{\ec}{\end{center}}
\newcommand{\ben}{\begin{enumerate}}
\newcommand{\een}{\end{enumerate}}

\newcommand{\dslash}{{\partial\kern-0.55em/}}

\def \muslash {\mu \kern-0.5em\slash}

\renewcommand{\O}{{\cal O}}

\newcommand{\skipover}[1]{}

\def\phm{\phantom{-}}

\newcommand{\keV}{{\rm keV}} 
\newcommand{\MeV}{{\rm MeV}} 
 
\newcommand{\pslash}{{p\kern-0.45em/}}
\newcommand{\kslash}{{k\kern-0.5em/}}

\newcommand{\thrS}{\ensuremath{{\bf 3}_S}}
\newcommand{\thrbar}{\ensuremath{\bar{\bf 3}}}
\newcommand{\thrbarA}{\ensuremath{\bar{\bf 3}_A}}
\newcommand{\six}{\ensuremath{{\bf 6}}}
\newcommand{\sixS}{\ensuremath{{\bf 6}_S}}

\newcommand{\Lag}{{\cal L}}
\newcommand{\onesc}{\ensuremath{({\rm 1SC})^3}}
\newcommand{\csl}{\ensuremath{({\rm CSL})^3}}
\newcommand{\twocsl}{2SC+CSLs}
\newcommand{\twoonesc}{2SC+1SCs}

\usepackage[normalem]{ulem}  

\begin{document}

\title{\bf 
Single-flavour and two-flavour pairing in 
three-flavour quark matter
}

\author{Mark G. Alford}
\affiliation{Department of Physics, Washington University\\ 
  St Louis, MO, 63130, USA}
\author{Greig A. Cowan}
\affiliation{Department of Physics and Astronomy, University of Glasgow\\
  Glasgow, G12 8QQ, UK}


\date{2005-Dec-8}


\begin{abstract}
We study single-flavour quark pairing (``self-pairing'')
in colour-superconducting phases of quark matter,
paying particular attention to the difference between
scenarios where all three flavours undergo single-flavour pairing,
and scenarios where two flavours pair with each other
(``2SC'' pairing) and the remaining flavour self-pairs.
We perform our calculations in the mean field approximation
using a pointlike four-fermion interaction based on 
single gluon exchange.
We confirm the result from previous weakly-coupled-QCD 
calculations, that when all three flavours self-pair
the favored channel for each is colour-spin-locked (CSL) 
pseudoisotropic pairing.
However, we find that when the up and down quarks undergo 2SC pairing, 
they induce a colour chemical potential that disfavors the CSL phase.
The strange quarks then self-pair in a ``polar'' channel that
breaks rotational invariance, although the CSL phase may survive
in a narrow range of densities.
\end{abstract}

\pacs{12.38.-t,25.75.Nq}
\keywords{quark matter; colour superconductivity; 
2SC; CSL; neutrality; free energy}
\preprint{GUTPA ???}

\maketitle

\renewcommand{\thepage}{\arabic{page}}

\section{INTRODUCTION}
\label{sec:intro}

Matter at high density and sufficiently low temperature is expected to
form a colour superconducting condensate of quark pairs
\cite{BarroisPhD,Bailin:1983bm,ARW2,Rapp:1997zu} (for reviews, see
Ref.~\cite{Rajagopal:2000wf,alford_review,Alford:2002ng,Nardulli:2002ma,Schafer:2003vz}).
Such a phase of matter may exist in the cores of
compact stars \cite{Glendenning} or be created during low energy
heavy-ion collisions \cite{Senger:2005pu}. At high density, quarks will fill
up the available energy states to form a Fermi surface. Since 
two quarks in the antisymmetric colour antitriplet channel
experience an attractive interaction, we expect a
BCS pairing instability \cite{bcs1}
at the Fermi surface. The instability is
resolved via the opening of a energy gap in the quasiparticle
spectrum. The quark pairing breaks the $SU(3)_{\text{colour}}$ gauge
symmetry of QCD, justifying the name colour superconductor.

The strong interaction is most attractive between quarks in
the colour-antisymmetric spin-0 channel, but Fermi-Dirac statistics 
then require flavour-antisymmetric pairing, involving two different
flavours. Well-studied candidates include the 2SC
\cite{ARW2,Rapp:1997zu,Buballa:2002xv} and CFL
\cite{ARW3,Alford:1999pa,Buballa:2001gj} phases of quark matter.
At asymptotically high
density, the favored phase of quark matter is the CFL phase.

However, the cores of compact stars are not asymptotically dense and
it becomes necessary to consider real-world effects such as a
non-zero strange quark mass, electrical and colour charge neutrality
and $\be$-equilibrium. Taken together, these constraints act to
separate the Fermi surfaces of the different quark flavours.
This means that as density decreases from the asymptotic regime
it becomes harder and harder to maintain the CFL condensate,
and we expect transitions to other phases.
In such a context it is natural to look for single-flavour pairing patterns,
and many different ones have been found
\cite{IwaIwa,TS1flav,BHO,Alford:2002kj,Schmitt:2004hg}. 
Calculations in weak-coupling QCD
indicate that for a single isolated flavour
the colour-spin-locked (CSL) phase has the
lowest free energy. However, in
the realistic context of quark matter in neutron
stars there are three flavours present. In this paper
we explore the single-flavour phases in such a context,
using a Nambu-Jona-Lasinio model. We find that if
the up and down quarks undergo two-flavour (2SC) pairing
then this typically induces a non-zero colour chemical potential
that disfavors CSL pairing for the remaining flavour,
which then self-pairs in a different ``transverse polar'' channel instead,
which we will simply label as ``1SC''.

The 1SC/polar channel was studied in the NJL context 
in Ref.~\cite{Alford:2002kj}, where we surveyed the pairing patterns 
that factorize into a product of colour, flavour, and Dirac (spin)
structures (this ansatz excludes the CSL phase, where colour
and spin are intertwined).
We used an NJL model with various plausible interactions,
and found that for a single flavour the most attractive channel
was the (\thrbarA,\thrS,1,$+$)($C\ga_3$) channel, where the
notation translates as (colour antisymmetric $\thrbar$,
flavour symmetric $\six$, spin 1, parity even), with 
Dirac structure $C\ga_3$. Pairing between quarks of the same
flavour in this channel was found to have a gap parameter of $\sim
1-10$ MeV. Unlike the CSL phase, this condensate breaks rotational
symmetry.
In weak-coupling QCD calculations this is called the
``transverse polar'' phase \cite{Schmitt:2004et}.

\begin{figure}[t]
 \includegraphics[width=0.5\textwidth]{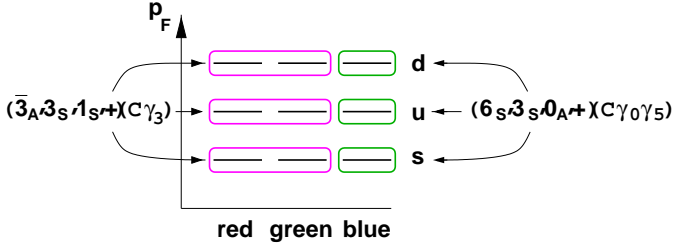}
 \caption[Pictorial representation of 1SC pairing in neutral quark matter.]{
Pictorial representation of simple single-flavour pairing in
neutral quark matter. This will be referred to as the \onesc\ phase in
the text. The requirement of electric neutrality and a nonzero strange quark
mass forces the Fermi momenta of the three flavours apart.
Two colours of each flavour form (\thrbarA,\thrS,1,$+$)($C\ga_3$)
Cooper pairs (1SC$u$, 1SC$d$ and 1SC$s$). The third colour of each
flavour forms (\sixS,\thrS,0,$+$)($C\ga_0\ga_5$) pairs.
}
\label{fig:1sc_levels}
\end{figure}

\begin{figure}[t]
\includegraphics[width=0.5\textwidth]{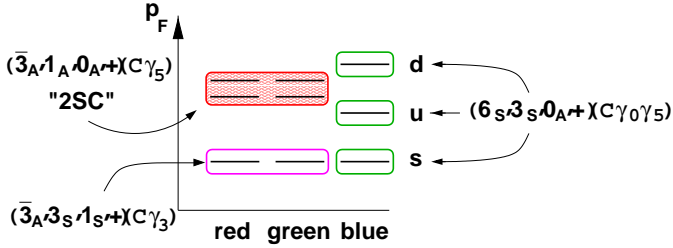}
\caption[Pictorial representation of 2SC and 1SC pairing 
in neutral quark matter.]{Pictorial representation of 
simultaneous 2SC and 1SC
pairing in neutral quark matter. This will be referred to as the
\twoonesc\ phase in the text. This will only occur if the
condensation energy of the 2SC pairing is strong enough to offset
the cost of dragging the red and green $u$ and $d$ Fermi momenta
away from the values dictated by electrical neutrality and the
strange quark mass, to a common value.
\label{fig:2sc_levels}}
\end{figure}

\begin{figure}[t]
\includegraphics[width=0.25\textwidth]{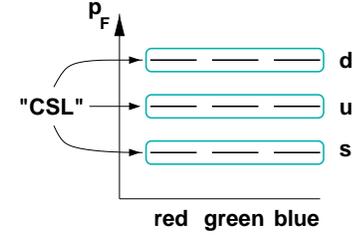}
\caption[Pictorial representation of CSL pairing
in neutral quark matter.]{Pictorial representation of CSL pairing in
  neutral quark matter. This will be referred to as the \csl\ phase in
  the text. It is composed of up, down and strange quark CSL
  condensates, labelled CSL$u$, CSL$d$ and CSL$s$ respectively. The
  requirement of electric neutrality and a nonzero strange quark mass
  forces the Fermi momenta of the three flavours apart. The red, green
  and blue colours of each flavour pair in a colour-antisymmetric
  channel.
  \label{fig:CSL_levels}}
\end{figure}

\begin{figure}[t]
\includegraphics[width=0.5\textwidth]{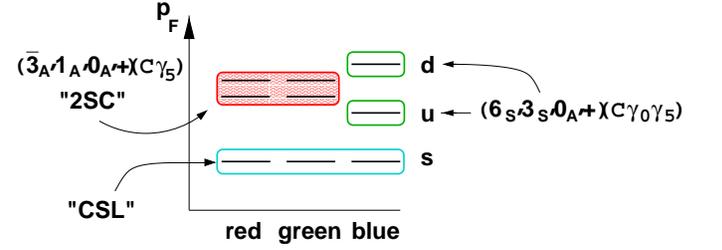}
\caption[Pictorial representation of 2SC and CSL pairing 
in neutral quark matter.]{Pictorial representation of 2SC and CSL
pairing in neutral quark matter. This will be referred to as the
\twocsl\ phase in the text. Colour neutrality will create a small
$\O(\De^2/\mu)$ splitting between the Fermi momenta of the blue $s$
quarks and $s$ quarks of the other two colours. If this splitting is
sufficiently large it will prevent the CSL$s$ condensate from forming.
\label{fig:2SC_CSL_levels}}
\end{figure}

In this paper we do not attempt a comprehensive study of all the
phases that have been suggested for quark matter. Our purpose is to
show that the presence of 2SC pairing can affect the single-flavour
pairing of the remaining flavour, so we restrict our study to
polar and CSL pairing of individual flavours, and 2SC pairing
of the up and down quarks. We do not take into account the
CFL phase, nor its gapless variant gCFL \cite{Alford:2003fq,Alford:2004hz}, 
nor crystalline (LOFF) pairing \cite{OurLOFF,Bowers:2002xr,Bowers_thesis},
and so on. 
Our study is relevant to regions of the phase diagram where
three flavours of quark are present, but there is either
no cross-flavour pairing, or only pairing between the
up and down quarks (``2SC+s'' in the the nomenclature of
Ref.~\cite{Alford:1999pa}). Such regions have been found to exist
in studies of the QCD phase diagram using NJL models.
For example Ref.~\cite{Ruster:2005jc} finds a 2SC phase with strange quarks
for stronger diquark coupling ($G_D=G_S$)
at $T=0$, over the chemical potential range 
$398~\MeV \lesssim \mu \lesssim 412~\MeV$
(see their Figs.~2 and 6). Ref.~\cite{Abuki:2005ms} also
finds such a phase (see their Fig.~7). It is also interesting to
note that there is some chance our considerations might be relevant
to the early life of the star, when $T\gtrsim 1~\MeV$,
neutrinos are trapped, and the resultant lepton number chemical
potential can favor a 2SC phase with strange quarks \cite{Steiner:2002gx}.
However some studies only find up and down quarks at this time 
\cite{Ruster:2005ib}. Moreover,
single flavour phases will only be relevant if they
have large enough gaps to survive at these temperatures. We find that their
critical temperatures are of order 1~\MeV (see Fig.~\ref{fig:1SC_CSL_gaps_T}),
but this result is sensitive to the cutoff 
(see end of Section \ref{sec:results}) so it remains an open
question whether they can survive during the era of neutrino trapping.

The paper has the following structure. Section \ref{sec:1SC/CSL}
summarizes the 1SC and CSL pairing patterns.
Sections \ref{sec:outline} and \ref{sec:model} discuss the
model and approximations used to calculate the free energy of the
colour superconducting phases that are of interest here.
In all cases, electrical
and colour neutrality constraints are imposed. Section \ref{sec:results}
shows the behaviour of the 1SC and CSL gap parameters as functions of
$\mu$ and $T$. The gaps are then used in a free energy calculation,
the results of which are presented in Section \ref{sec:phase_diags} as
a sequence of ($\mu,T$) phase diagrams for neutral three flavour quark
matter for different values of $M_s$. Section \ref{sec:csl_summ}
presents conclusions and directions for further research.

\clearpage

\section{1SC and CSL pairing}
\label{sec:1SC/CSL}

The forms of zero-temperature pairing that will be compared
in this paper are shown in Figs.~\ref{fig:1sc_levels},
\ref{fig:2sc_levels}, \ref{fig:CSL_levels}, \ref{fig:2SC_CSL_levels}.
In the ``\onesc''
phase (Fig.~\ref{fig:1sc_levels}), the Fermi momenta of the three 
flavours are so different that only single-flavour pairing is possible.
Two colours of each flavour form (\thrbarA,\thrS,1,$+$)($C\ga_3$)
Cooper pairs (1SC$u$, 1SC$d$ and 1SC$s$).
The third colour of each
flavour forms (\sixS,\thrS,0,$+$)($C\ga_0\ga_5$) pairs. 
In the \twoonesc\ phase (Fig.~\ref{fig:2sc_levels}), the up and down
quarks are able to pair with each other in the 2SC channel (which
only involves two colours), but the strange quarks undergo single-flavour
pairing. Again, the third colour of each
flavour forms (\sixS,\thrS,0,$+$)($C\ga_0\ga_5$) pairs. 

Figures \ref{fig:CSL_levels} and \ref{fig:2SC_CSL_levels} are the
equivalents of Figures \ref{fig:1sc_levels} and
\ref{fig:2sc_levels}, where the 1SC pairing of individual
flavours is replaced by CSL pairing.


We now give the explicit form of the pairing patterns mentioned so far.
The basic two-colour single-flavour spin-1 1SC pairing
(the ``polar phase'' in Ref.~\cite{TS1flav}), described here as
(\thrbarA,\thrS,1,$+$)($C\ga_3$), is
\begin{equation}
  \label{eq:1SC_ansatz}
  \mathbf{\De}^{ab,ij}_{\al\be} 
  = \De_{1SC}(C\ga_3)^{ab}(\la_2)_{\al\be}\de^{ij} \ .
\end{equation}
The colour indices are $\al,\be$, and $\la_2$ is an
antisymmetric Gell-Mann matrix in colour space. The Dirac indices
are $a,b$, and $C=i\ga_0\ga_2$ is the usual charge conjugation matrix.
The flavour indices $i,j$ are included for completeness: in this single-flavour
channel $i=j=1$.
This condensate breaks rotational invariance, as the explicit
occurrence of the $\ga_3$ matrix indicates. 
Only two colours (typically taken as red and green) of quark are
involved in the pairing meaning that the remaining (blue) quark is
required to find an alternative channel in which to pair. The
$SU(3)_{\rm colour}$ symmetry is broken down to an $SU(2)$
subgroup.

The single-colour single-flavour (\sixS,\thrS,0,$+$)($C\ga_0\ga_5$)
pairing is
\begin{equation}
  \mathbf{\De}^{ab,ij}_{\al\be} 
  = \De_{05}(C\ga_0\ga_5)^{ab}\de_{\al\be}\de^{ij} \ .
\end{equation}
This channel is attractive in an NJL model with
gluon-exchange-type interaction, but
the pairing disappears as the quark masses go to zero \cite{Alford:2002kj}.
This means it may be somewhat suppressed for up and down quarks,
depending on how much of their vacuum constituent mass survives
at the relevant densities \cite{BHO,Huang:2001yw,Blaschke:2003cv}.
Our earlier NJL study estimated the pairing gap in the $C\ga_0\ga_5$ channel 
to be of the order 1-100 eV. This is
much smaller than the typical gaps in the 2SC, 1SC, and
CSL channels, so when we perform free energy calculations we will
neglect the (\sixS,\thrS,0,$+$)($C\ga_0\ga_5$) pairing, treating
those quarks as unpaired.

CSL pairing is
\begin{equation}
  \label{eq:CSL_ansatz}
  \mathbf{\De}^{ab,ij}_{\al\be} 
   = \De_{\rm CSL}\,(C\ga_A)^{ab}\,\la^A_{\al\be}\,\de^{ij},
\end{equation}
where we sum over $A=1,2,3$, $\ga_A = (\ga_1,\ga_2,\ga_3)$ and $\la_A =
(\la_7,\la_5,\la_2)$ which are the antisymmetric 
Gell-Mann matrices in colour space. 
The symmetry breaking associated with the CSL phase is
\begin{equation}
  \label{eq:CSL_symm_break}
  SU(3)_{\rm colour} \times SO(3)_J \times U(1)_B \rightarrow 
  SO(3)_{{\rm colour}+J} \times \mathbb{Z}_2,
\end{equation}
where $SU(3)_{\rm colour}$ is gauged, $SO(3)_J$ is the 
rotation group and
$U(1)_B$ is baryon number.
The unbroken rotational symmetry
$SO(3)_{{\rm colour}+J}$ corresponds to a locking between the colour and
spin degrees of freedom. This can be thought of as associating a
direction in configuration space with a linear combination of
directions in colour space, i.e. red and green quarks pair in the
$z$-direction, red and blue quarks in the $y$-direction and green and
blue quarks in the $x$-direction. Therefore, unlike the polar phase,
CSL pairing remains isotropic. The locking is similar to that which is
found in the CFL phase between colour and flavour. The CSL phase is
the colour superconducting equivalent of the $B$-phase of superfluid
helium-3 \cite{helium3}.

In general, one would expect that the CSL condensate could involve
a linear combination of $C \ga_i$ and $C \si_{0i}$ with a
colour structure that is correlated with the spatial direction.
However, Table 1 in Ref.~\cite{Alford:2002kj} shows that the colour
antisymmetric $C\sigma_{03}$ channel gives no contribution when using
the mean-field approximation within an NJL model with a full gluon
interaction vertex.  This is precisely the interaction that we use,
so only the $C \ga_i$ term is included in the CSL ansatz of
Equation \eqn{eq:CSL_ansatz}.

We can calculate the binding energy of the
CSL channel, following Ref.~\cite{Alford:2002kj}, and compare it with the
results for the polar 1SC channel, taken from the same
reference. The result is shown in Table \ref{tab:MF_CSL},
where for a given condensate of gap parameter $\De$
the binding energy via interaction $I$ is given by
$E = - S_I \De^2$, so larger values of $S$ imply stronger binding.
$I$ can be ``inst'' (instanton interaction), ``mag'' (magnetic
gluon interaction), or ``elec'' (electric gluon interaction).
The table shows that the CSL condensate is more tightly bound
than the polar condensate, leading one to expect that, other
things being equal, CSL pairing will be favored over polar pairing.

\begin{table*}
\newcommand{\st}{\rule[-2ex]{0em}{4ex}} 
\setlength{\tabcolsep}{0.25em}
\newlength{\instwid}
\newlength{\gluonwid}
\newlength{\magwid}
\settowidth{\instwid}{instanton} 
\settowidth{\gluonwid}{x+$S_{\rm mag}$x}
\settowidth{\magwid}{mag.~only}
\begin{tabular}{ccccccccccccc}
\hline\hline
\st phase & colour & flavour & $j$ & parity &
\multicolumn{2}{c}{Dirac} &  
\parbox{4em}{\bc BCS\\[-0.5ex] enhance-\\[-0.5ex] ment\ec}
 & \parbox{\instwid}{\bc ~ \\[-0.2ex] $S_{\rm inst}$ \ec}
 & \parbox{\gluonwid}{
      \bc $S_{\rm elec}$\\[-0.2ex]+$S_{\rm mag}$ \ec}
 & \parbox{\magwid}{\bc ~ \\[-0.2ex] $S_{\rm mag}$ \ec } \\[-1ex]
\hline
\st CSL & \thrbarA & \thrS & $1_S$ & $+$& $C\ga_i$ & LR & $\O(1)$ 
  & 0 & $+96$ & $+48$ \\
\st ``Polar'' (1SC) & \thrbarA & \thrS & $1_S$ & $+$& $C\ga_3$ & LR & $\O(1)$ 
  & 0 & $+32$ & $+16$  \\
\hline\hline
\end{tabular}
\caption[Binding strengths of CSL channel in NJL models
in the mean-field approximation.]
{Binding strengths of CSL (1st row) and 1SC (2nd row) channels in an NJL model
in the mean-field approximation. The 1SC results are taken from  
Table 1 in \cite{Alford:2002kj}. We see that CSL is more strongly bound.
\label{tab:MF_CSL}
} 
\end{table*}

\section{Outline of Calculation}
\label{sec:outline}

The aim of this paper is to map out the $(\mu,T)$ phase diagram 
for the 2SC, unpaired, \onesc, \csl, \twoonesc, and \twocsl\ phases of
neutral three flavour quark matter at densities relevant for compact
stars ($\mu\sim 500$ MeV), for different values of $M_s$.
The phase diagram is obtained by calculating the free energy
of each phase, imposing neutrality conditions and
equilibration under the weak interactions, and minimizing
the free energy with respect to the pairing strengths $\De_\rho$
in the various phases. In section \ref{sec:model} we will
discuss the NJL model within which the free energy is calculated.
In this section we assume that the free energy is known.

To obtain neutral matter we introduce chemical potentials
coupled to the gauged charges. 
(In full QCD the time-components of the gauge fields play this
role \cite{Gerhold:2003js,Dietrich:2003nu}, and there are no
gluonic contributions to the charge density \cite{Gerhold:2004sk}.)
The electron chemical potential $\mu_e$ couples to negative electric
charge. To impose colour neutrality it is sufficient to couple
chemical potentials $\mu_3$ and $\mu_8$ to the Cartan sub-algebra
of the colour gauge group, generated by
$T_3=\mbox{diag}(1/2,-1/2,0)$ and
$T_8=\mbox{diag}(1/3,1/3,-2/3)$ \cite{Buballa:2005bv}.

The charge densities are related to the chemical potentials via the
derivatives of the free energy, so the neutrality constraints are
\beq
\ba{rcl}
\label{neutrality}
 \dsp n_8 \equiv -\frac{\partial\Omega}{\partial\mu_8} &=& 0,\\[3ex]
 \dsp n_3 \equiv -\frac{\partial\Omega}{\partial\mu_3} &=& 0,\\[3ex]
 \dsp n_Q \equiv \phm\frac{\partial\Omega}{\partial\mu_e} &=& 0.
\ea
\eeq

The weak interaction violates flavour conservation but conserves
baryon number, allowing the chemical potentials of the individual
quark species to be written in terms of 
$\mu_e$, $\mu_3$, $\mu_8$ and the quark number chemical potential $\mu$,
\cite{Alford:2004hz}
\beq
\ba{rcl}
 \mu_{ru} &=& \mu - \frac{2}{3}\mu_e + \frac{1}{2}\mu_3 + \frac{1}{3}\mu_8,\\
 \mu_{gu} &=& \mu - \frac{2}{3}\mu_e - \frac{1}{2}\mu_3 + \frac{1}{3}\mu_8,\\
 \mu_{bu} &=& \mu - \frac{2}{3}\mu_e - \frac{2}{3}\mu_8,\\
 \mu_{rd} &=& \mu + \frac{1}{3}\mu_e + \frac{1}{2}\mu_3 + \frac{1}{3}\mu_8,\\
 \mu_{gd} &=& \mu + \frac{1}{3}\mu_e - \frac{1}{2}\mu_3 + \frac{1}{3}\mu_8,\\
 \mu_{bd} &=& \mu + \frac{1}{3}\mu_e - \frac{2}{3}\mu_8,\\
 \mu_{rs} &=& \mu + \frac{1}{3}\mu_e + \frac{1}{2}\mu_3 + \frac{1}{3}\mu_8 
 - \frac{M_s^2}{2\mu} - \frac{M_s^4}{8\mu^3},\\
 \mu_{gs} &=& \mu + \frac{1}{3}\mu_e - \frac{1}{2}\mu_3 + \frac{1}{3}\mu_8 
 - \frac{M_s^2}{2\mu} - \frac{M_s^4}{8\mu^3},\\
 \mu_{bs} &=& \mu + \frac{1}{3}\mu_e - \frac{2}{3}\mu_8 
 - \frac{M_s^2}{2\mu} - \frac{M_s^4}{8\mu^3}.
\ea
  \label{eq:chem_pots}
\eeq
The strange quark mass is included to leading order as
a term in the strange quark chemical potential 
\cite{Bedaque:2001je,Alford:2004hz}.

The neutrality constraints will always be implemented locally,
i.e.~no mixed phases of quark matter and nuclear matter
will be discussed \cite{Neumann:2002jm,Reddy:2004my}.

For each condensate $\rho$, the gap equation is found analytically by
differentiating the free energy with respect to the gap parameter of
that condensate,
\begin{equation}
  \frac{\partial\Om_{\rm total}}{\partial\De_\rho} = 0. 
\end{equation}
Finding the free energy of a phase therefore corresponds to finding
a stationary point of $\Om$ in the parameter space of chemical
potentials and gap parameters.
If we work in the neutral subspace, fixing $\mu_e,\mu_3,\mu_8$
as functions of the $\De_\rho$, then the relevant stationary point
will always be a minimum with respect to the gap parameters.


We can now anticipate some simplifications that will streamline
our calculation.

Firstly, as is clear from Figs.~\ref{fig:1sc_levels},
\ref{fig:2sc_levels}, \ref{fig:CSL_levels}, \ref{fig:2SC_CSL_levels},
we will always be concerned with 
colour superconducting condensates that are at least invariant under
the $SU(2)$ colour subgroup that rotates red and green
 quarks into each other. This means that $\mu_3$ will always be zero.

Secondly, it is never necessary to solve the full set of coupled
gap and neutrality equations. The single-flavour pairing is so weak
that its back-reaction on the chemical potentials via the neutrality
condition can be neglected. Thus for the \twoonesc\ and \twocsl\
phases one can first treat the strange quarks as free, determine
$\mu_e$ and $\mu_8$ by solving for 2SC pairing, and then use
those values in the gap equations for the strange quark pairing.
Similarly, for the \onesc\ and \csl\ phases one can use the values
of $\mu_e$ and $\mu_8$ that would ensure neutrality in totally 
unpaired quark matter (in the case of $\mu_8$ this value is zero).

Actually, the only phase where the exact value of $\mu_8$ is important 
is the \twocsl\ phase, where the nonzero $\mu_8$
puts a stress on the CSLs pairing.
Following the standard argument \cite{Alford:2003fq}, we expect that
CSL pairing can occur as long as the cost of breaking a CSL pair
is greater than the gain in energy from converting a blue quark
to a red/green one, i.e.,
\begin{equation}
  \label{eq:CSL_pair_cond}
  |\mu_8| < 2\De_{\rm CSL}. 
\end{equation}
The phase diagrams to
be presented in section \ref{sec:phase_diags} will show this
behaviour.

\section{Model and approximations}
\label{sec:model}

The NJL-type
Lagrangian that will be used in the following calculations is,
\begin{eqnarray}
  \label{eq:L_NJL2}
  \Lag_{\rm NJL} & = & \Lag_{\rm kin} + \Lag_{\rm int}\\
  & = & \bar{\psi}( i \dslash - \muslash) \psi + \Lag_{\rm int},
\end{eqnarray}
where $\muslash = \mu\ga_0$ is a diagonal matrix in colour-flavour
space, with entries given by \eqn{eq:chem_pots}. 
The up and down quarks are treated as massless, and
the strange quark mass is only taken into account 
to leading order, as an effective chemical potential for strangeness.
The interaction we use is abstracted
from single-gluon exchange by replacing the gluon propagator with
a four-fermion coupling constant,
\begin{equation}
  \label{eq:L_int1}
  \Lag_{\rm int} = -\frac{3}{8}\,G
  (\bar{\psi}\Gamma^a_\mu\psi)(\bar{\psi}\Gamma_a^\mu\psi),
\end{equation}
where
\begin{equation}
  \label{eq:vertex}
  \Gamma^a_\mu = \gamma_\mu \lambda^a,
\end{equation}
where the $\lambda^a$'s are the standard Gell-Mann matrices
($a=1\ldots 8$), giving the interaction the quantum numbers of single
gluon exchange. 

In order to study diquark pairing and colour superconductivity, we
Fierz transform the interaction into the form
\begin{equation}
  \label{eq:L_int2}
  \Lag_{\rm int} = G \sum_\rho \alpha'_\rho 
  (\bar{\psi}\Gamma^{qq}_\rho\bar{\psi}^T)
  (\psi^T\bar{\Gamma}^{qq}_\rho\psi),
\end{equation}
where $\rho$ labels the different diquark pairing channels,
$\Ga^{qq}_\rho$ specifies the
form of the diquark pairing in channel $\rho$ and
$\bar{\Gamma}^{qq}_\rho$ is defined by
\begin{equation}
  \Ga^{qq}_\rho = \ga_0(\bar{\Ga}^{qq}_\rho)^\dagger\ga_0^T.
\end{equation}

Note that we sum up the terms for all the different pairing channels,
even though no single Fierzing of the original interaction would
give the sum of all of them. This is legitimate because there
are no cross-terms between different channels in the binding energy.

With the interaction expressed as in Eq.~\ref{eq:L_int2} it is
straightforward to do a mean field calculation of the free energy is
now in a suitable form to allow the free energy to be calculated. This
is done by performing a bosonisation of the Fierz transformed
Lagrangian via a Hubbard-Stratonovich transformation
\cite{Kikkawa:1976fe,Ebert:1991pz,Reinhardt:1989rw,Berges:1998rc}.
The general expression for the
thermodynamic potential of a colour superconducting condensate is
\begin{equation}
  \label{eq:thermo_pot_eqn}
  \Omega = -T \sum_n \int 
  \frac{d^3p}{(2\pi)^3}\,\frac{1}{2}{\rm Tr\, ln}
  \left(\frac{S^{-1}(i\omega_n,\vec p)}{T}\right) 
  + \frac{\De_\rho\De_\rho}{4\alpha^\prime G},
\end{equation}
where $S^{-1}(i\omega_n,p)$
is the full inverse quark propagator.
$\omega_n=(2n-1)\pi T$ are the Matsubara frequencies and the trace
is over colour, flavour, and spinor indices.
The Matsubara summation is performed using the identity
\begin{equation}
  \label{eq:matsubara1}
  T\sum_n\ln\left(\frac{\omega_n^2 + \eps_a(\vec p)^2)}{T^2}\right)
= |\eps_a(\vec p)| + 2T\ln(1 + e^{-|\eps_a(\vec p)|/T}).
\end{equation}
The functions $\eps_a(p)$ are the dispersion relations for the
fermionic quasiparticles.
They are not explicitly $T$-dependent, but
they depend upon the gap and quark chemical potentials which are
$T$-dependent. 

For convenience, we use neutral unpaired quark
matter as the zero of free energy, so we subtract it from the
value computed above. This gives a physically meaningful quantity
without ultraviolet divergences:
\beq
\ba{r@{~}c@{~}l}
  \label{eq:FE_Tneq0}
  \Om =  &-&\dsp \int \frac{d^3p}{(2\pi)^3}\,\frac{1}{2}
  \sum_a \left(|\eps_a(\vec p)| 
     + 2T\ln(1 + e^{-|\eps_a(\vec p)|/T})\right)\\[2ex]
  &+&\dsp \frac{\De_\rho\De_\rho}{4\alpha^\prime G}
   - \frac{1}{12\pi^2}\left(\mu_e^4 + 2\pi^2T^2\mu_e^2 
     + \frac{7\pi^4}{15}T^4\right) \\[2ex]
  &-&\Om_{\rm free}(\mu,M_s,\mu^{\rm unp}_e,T) 
\ea
\eeq
where the electron contribution to the free energy has been included.
The electron chemical potential in
neutral unpaired quark matter is
\begin{equation}
  \label{eq:neutral_sols}
  \mu_e^{unp} = \frac{M_s^2}{4\mu} \ .
\end{equation}
The quasiquark dispersion relations $\eps_a(\vec p)$
are the values of the energy at
which the propagator diverges, i.e.
\begin{equation}
  \label{eq:detS}
  \mathrm{det}\,S^{-1}(\eps_a(\vec p),\vec p) = 0. 
\end{equation}
The calculation of $\Om$ then depends upon the calculation of the
determinant of the matrix $S^{-1}$. This matrix can be block-diagonalized
in the colour-flavour space, with one block for each independent pairing
channel. For each colour and flavour there are eight dispersion relations
(four Dirac branches including left-handed and right-handed, 
particle and antiparticle, doubled by the Nambu-Gorkov formalism).
In calculating the free energy of quark matter with three flavours
and three colours the index $a$ in \eqn{eq:detS} will therefore
vary from 1 to 72.

\subsection{Dispersion relations}
\label{sec:disp_rels}

For each unpaired colour and flavour of quark, there two branches
of the dispersion relation (particle and antiparticle) each with
multiplicity 4:
\beq
E^\pm_{\rm free} = p\pm\mu
\eeq

For each condensate we now give the dispersion relation
of the quasiquarks (with their multiplicity)
and the binding energy parameter $\al'$.

\subsubsection{2SC}
For 2SC pairing of the red and green quarks
of up and down flavour, the dispersion relations are
\beq
  E_{\rm 2SC} =
  \sqrt{(p \pm \bar{\mu})^2 + \De^2} \pm \de\mu
\eeq
each with multiplicity 4 (spin and Nambu-Gorkov), where
\beq
\ba{rcl}
  \bar{\mu} &=& \dsp \frac{1}{2}(\mu_{ru} + \mu_{gd}) = \frac{1}{2}(\mu_{ru} + \mu_{gd})\\ 
  &=& \dsp \mu - \frac{\mu_e}{6} + \frac{\mu_8}{3},\\
  \de\mu &=& \dsp \frac{1}{2}(\mu_{gd} - \mu_{ru}) = \frac{1}{2}(\mu_{rd} - \mu_{gu}) 
  = \frac{\mu_e}{2}\ .
\ea
\eeq
The binding energy parameter is
\begin{equation}
  \al^\prime = \frac{1}{4}.
\end{equation}
If $\De_{2SC} < \de\mu$, then the 2SC condensate is in the gapless 2SC
state \cite{Shovkovy:2003uu,Huang:2003xd}.

\subsubsection{1SC}
For 1SC pairing of red and green quarks of some flavour $f$
(assumed to have common chemical potential $\bar\mu$),
the dispersion relations are
\begin{equation}
  E_{\rm 1SC}^2 = 
  p^2 + m^2 + \bar{\mu}^2 + \De^2
 \pm 2\sqrt{ \bar{\mu}^2(p^2+m^2) +  \De^2 \,p_3^2} \ ,
\end{equation}
each with multiplicity 4. The binding energy parameter is
\begin{equation}
  \al^\prime = \frac{1}{8}.
\end{equation}

\subsubsection{CSL}
For CSL pairing of three colours of a single massless flavour,
in the case $\mu_8=0$, we can obtain a simple closed-form expression for
the determinant of the inverse propagator,
\beq
\label{eq:CSLdet}
\ba{rcl}
 \multicolumn{3}{l}{\det\,S^{-1}(p_0,\,p) =} \\
  &&(\De^2 + \mu^2 - (p - p_0)^2)^2 \\
  &\times& (\De^2 + \mu^2 - (p + p_0)^2)^2 \\
  &\times& \Bigl(4 \De^4 + \mu^4 + (p^2 - p_0^2)^2 - 2 \mu^2 (p^2 + p_0^2)\\
  &&  + \De^2 (5 \mu^2 + (3 p - 5 p_0) (p + p_0))\Bigr)^2 \\
  &\times& \Bigl(4 \De^4 + \mu^4 + (p^2 - p_0^2)^2 - 2 \mu^2 (p^2 + p_0^2) \\ 
  &&  + \De^2 (5 \mu^2 + (3 p + 5 p_0) (p - p_0))\Bigr)^2 \ ,
\ea
\eeq
which has dimension 24, as expected.
From this the dispersion relations of the 24 quasiparticles can be obtained.
Note that there are 4 gapless modes (first line of the determinant),
and the remaining 20 are gapped. When we include a nonzero quark mass
all the quark modes are gapped, which is different from the
behaviour in weak-coupling QCD \cite{Schmitt:2004et}.
For a discussion of this point see Refs. \cite{Aguilera:2005tg,Schmitt:2005wg}.
In the general case $\mu_8\neq 0, m>0$,
the determinant can be readily obtained using a symbolic mathematics
program such as {\em Mathematica}, but is too complicated to display here.

The binding energy parameter is
\beq
  \al^\prime = \frac{1}{8}.  
\eeq

\section{Results}
\label{sec:results}

The following sections present the results of solving the gap and neutrality
equations and calculating the free energies for the 
\onesc, \csl, \twoonesc, and \twocsl\ condensates.
The momentum cutoff was fixed at $\Lambda=600$~MeV, with
the value of the coupling $G$ calibrated to give $\De_{2SC}(\mu=500,
M_s=0) = 31.1$~MeV, as in Ref.~\cite{Alford:2003fq}. This corresponds to
$\De_{\rm CFL}(\mu=500, M_s=0) = 25$ MeV. We also performed some calculations
at $\La=800~\MeV$ with the same calibration, to check for cutoff sensitivity.

If single-flavour pairing were ignored, the phase diagram would
contain a transition from unpaired quark matter 
to 2SC quark matter at $\mu = M_s^2/(2\mu)$. In principle this
critical value also depends on temperature, but at the
temperatures of interest to us ($T\lesssim 1~\MeV$) this
can be neglected.
Thinking of this as a background, we can then include the
possibility of 1SC and CSL single-flavour pairing, and we expect
to find that in the low-temperature parts of the unpaired region
there are regions of \onesc\ or \csl\ phases, and in the
low-temperature parts of the 2SC region there are regions of \twocsl\
or \twoonesc. This is indeed what we ultimately find.

As mentioned in section \ref{sec:model}, we can calculate $\mu_e$
and $\mu_8$ in the 2SC/unpaired background, before including the
effects of single flavour pairing. We show these results 
in the next subsection. Then we go on to solve the gap equations for
the various forms of single flavour pairing, and compare their
free energies.

\subsection{Chemical potentials with no single flavour pairing}
\label{sec:chem_pots}

\begin{figure}[t]
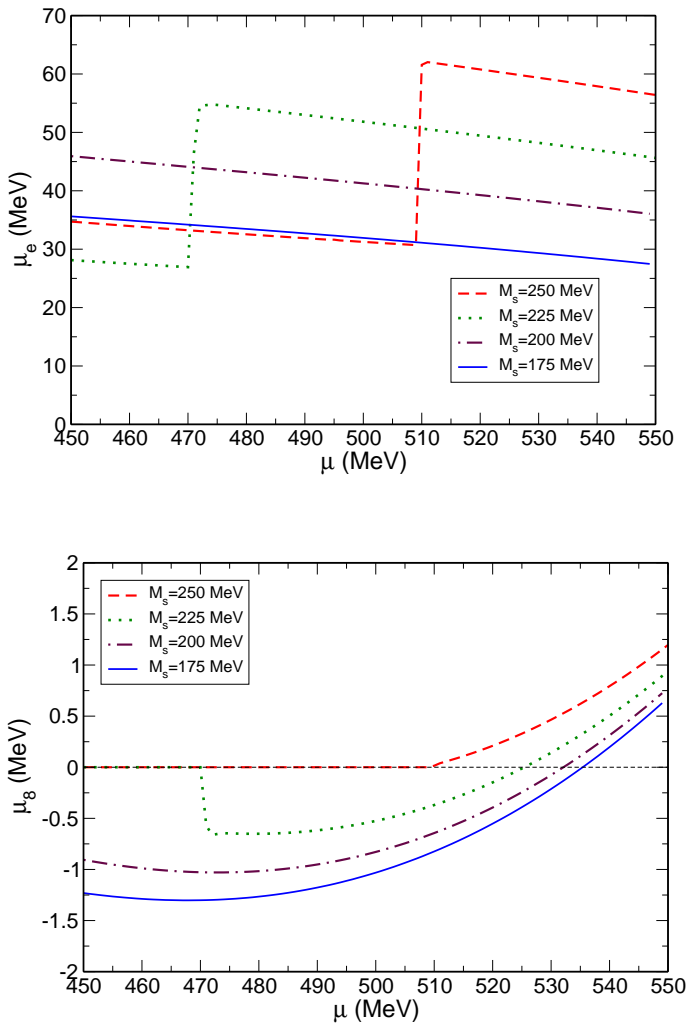

  \includegraphics[scale=0.36]{./figures/mue_mu_T0.03.eps}\\[1cm]
  \includegraphics[scale=0.36]{./figures/mu8_mu_T0.03.eps}
  \caption{\label{fig:mue_mu8_mu} 
For 2SC/unpaired quark matter, the variation of  $\mu_8$ 
with quark chemical potential at $T\approx 0$ and for various values 
of $M_s$. We see the transition from unpaired ($\mu_8=0$) at
low $\mu$ to 2SC at high $\mu$.
  }
\end{figure}

Figure \ref{fig:mue_mu8_mu} shows (for four different strange quark
masses) the values of $\mu_e$ and $\mu_8$ in neutral
unpaired/2SC matter at $T\approx 0$
over a range of quark chemical potential $450~\MeV < \mu < 550~\MeV$.
This required simultaneously solving the 2SC gap equation
and the two neutrality equations, and choosing the phase with
the lowest free energy at each value of $\mu$.

At $\mu\approx M_s^2/(2\mu)$
we see the transition from unpaired to 2SC, at which
$\mu_e$ jumps from about $M_s^2/(4 \mu)$ to about $M_s^2/(2\mu)$
\cite{AR-02}. In the unpaired phase,
$\mu_8$ is zero because nothing picks out any direction in colour space.
In the 2SC phase, however, the red and green light quarks pair, and
this means that a nonzero $\mu_8$ is required to maintain colour
neutrality. At moderate densities $\mu_8$ is negative
because the number of red and green quarks is enhanced
by the 2SC pairing (there is more phase space above the Fermi
surface than below it), and a negative value of $\mu_8$ restores
the colour balance by favoring blue quarks ($T_8={\rm diag}(1,1,-2)$).
As the quark chemical potential approaches the cutoff, however,
the situation is reversed: degrees of freedom above the cutoff
are excluded, so the condensate contains more red and green holes
below the Fermi surface than red and green quarks above it.
Thus $\mu_8$ turns around and starts to rise again as $\mu$ approaches
$\La$. Depending on the values of the strange quark mass
and the cutoff, $\mu_8$ may cross through zero and become positive
at some specific density, as seen in Fig.~\ref{fig:mue_mu8_mu}.
This is a very interesting phenomenon, because it leads to an
``island'' of \twocsl\ pairing in the phase diagram
(Fig.~\ref{fig:phase_diags}),
but we emphasize that it is sensitive to the value of the cutoff, and
if we work at $\La=800~\MeV$ then we find that $\mu_8$ 
only becomes positive at $\mu>\La$.

\subsection{Gap parameter plots}
\label{sec:gap_plots}

\begin{figure}[t]
  \includegraphics[scale=0.36]{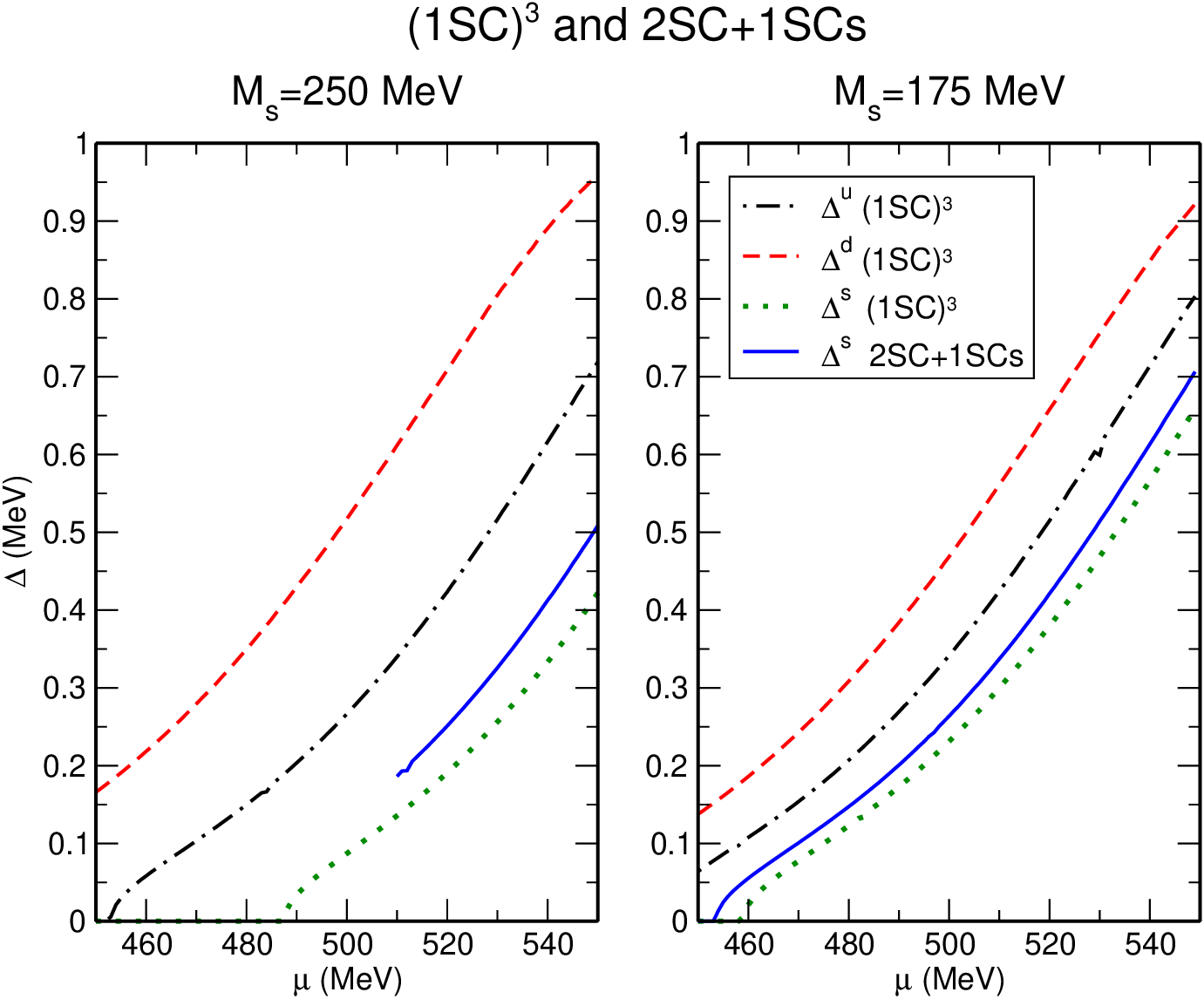}
  \\[0.5cm]
  \includegraphics[scale=0.36]{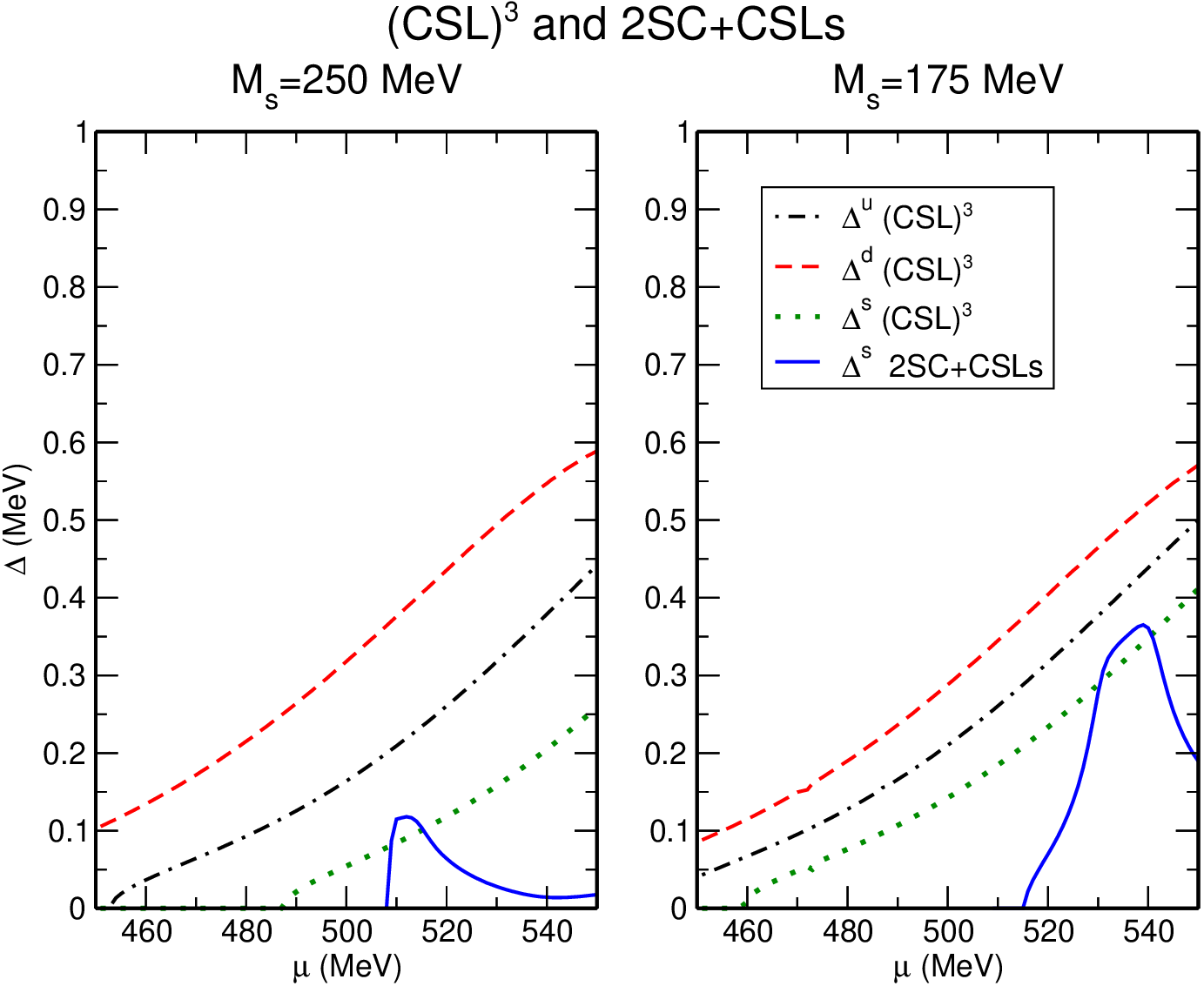}
  \caption{\label{fig:1SC_CSL_gaps_mu}
    Variation of the 1SC (upper) and CSL (lower) gaps with quark
    chemical potential at $T=0.03$ MeV and for $M_s=250,175$ MeV.
    Solid lines are for the strange quarks, in the case where up and down
quarks are undergoing 2SC pairing. The various broken lines are for
    the three flavours of quark in the case where all of them undergo
single-flavour pairing.}
\end{figure}

We now proceed to solve the gap equations for the single-flavour-pairing
phases.
Figure \ref{fig:1SC_CSL_gaps_mu} shows how the single-flavour pairing
gap parameters 
vary over the quark chemical potential range  $450~\MeV < \mu < 550~\MeV$.
In all cases, the gaps for the different flavours follow the same
pattern. 

For the purely single-flavour phases (\onesc\ and \csl)
there are three gap parameters, for up, down, and strange quarks
(red and green only, see Figs.~\ref{fig:1sc_levels},\ref{fig:CSL_levels}).
These are shown by
the various broken lines in Fig.~\ref{fig:1SC_CSL_gaps_mu}.
They exhibit the hierarchy,
\begin{equation}
  \label{eq:single_flav_order_mu}
  \De^d > \De^u > \De^s.
\end{equation}
This follows from the electrical-neutrality-induced ordering
of Fermi momenta in unpaired quark matter
\mbox{$p_F^d > p_F^u > p_F^s$}
and the fact that quarks with larger Fermi momentum have more phase space 
at their Fermi surface, and hence stronger pairing.
We also see that 1SC pairing gives a larger gap than CSL pairing,
but it will turn out that CSL has a larger condensate binding energy
(see Table \ref{tab:MF_CSL}) and so, other thing being equal, 
we expect CSL to be favoured over
1SC. This is confirmed by the free energy calculation and can be
seen by the presence of the CSL phase in the phase diagrams of section
\ref{sec:phase_diags}.

In the \twocsl\ and \twoonesc\ phases only the strange quarks
undergo single-flavour pairing, so there is only one
gap parameter, $\De^s$. This is shown
by the solid lines in Fig.~\ref{fig:1SC_CSL_gaps_mu}).
In the top left panel of the figure, the solid line 
(\twoonesc) ends at $\mu\approx 510~\MeV$, because the 2SC phase
ceases to be favored below that point. 
Notice that in general the value of $\De^s$ in the presence of
2SC pairing of the up and down quarks is slightly higher than its
value when all three flavours undergo single-flavour pairing.
This is because 2SC pairing pulls the up and down Fermi surfaces
together, which drags the strange Fermi momentum up a little
in order to maintain electrical neutrality. This increases the phase space
at the strange quark Fermi surface, boosting the gap by a small amount.

The gap parameters for the CSL$s$ condensate in the \twocsl\ phase 
show a non-monotonic behaviour (solid lines in lower two panels of
Fig.~\ref{fig:1SC_CSL_gaps_mu}). That is because CSL pairing
is very sensitive to the value of the $\mu_8$ chemical potential,
which tries to split apart the Fermi momentum of the
blue quarks from those of the red and green quarks, and the
background 2SC pairing induces a non-zero $\mu_8$ to
maintain colour neutrality. CSL pairing
is disfavored when its gap is less than a critical value
of order $\mu_8$. For the cutoff that we used, $\La=600~\MeV$,
$\mu_8$ goes through zero at $\mu\sim 530~\MeV$
(see Sect.~\ref{sec:chem_pots}), so the CSL gap is boosted there.
For a different cutoff, this would have occurred at a different density.

\subsection{Non-BCS temperature dependence}
\label{sec:non_BCS_temp}

\begin{figure}[t]
  \includegraphics[scale=0.36]{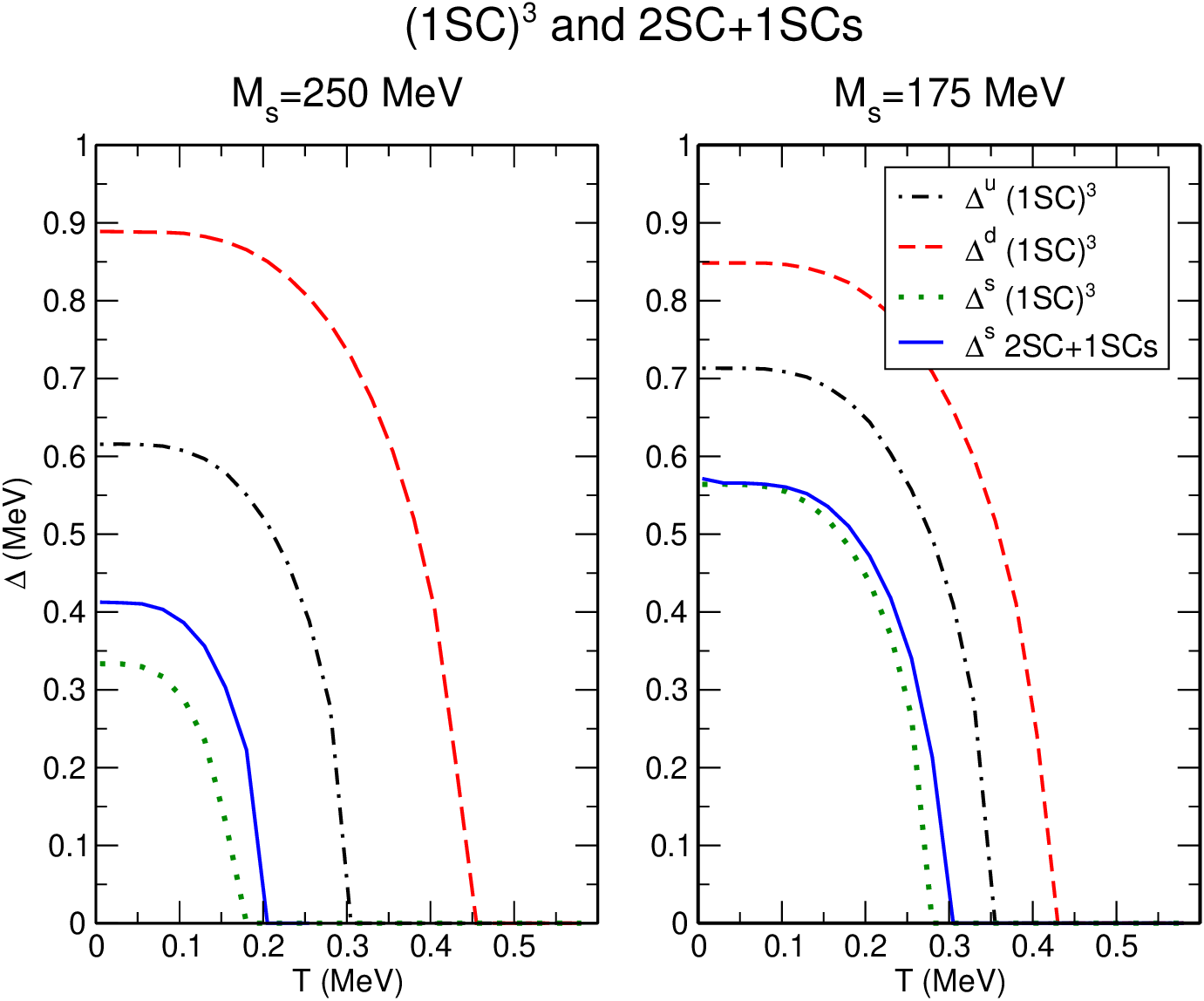}\\[0.5cm]
  \includegraphics[scale=0.36]{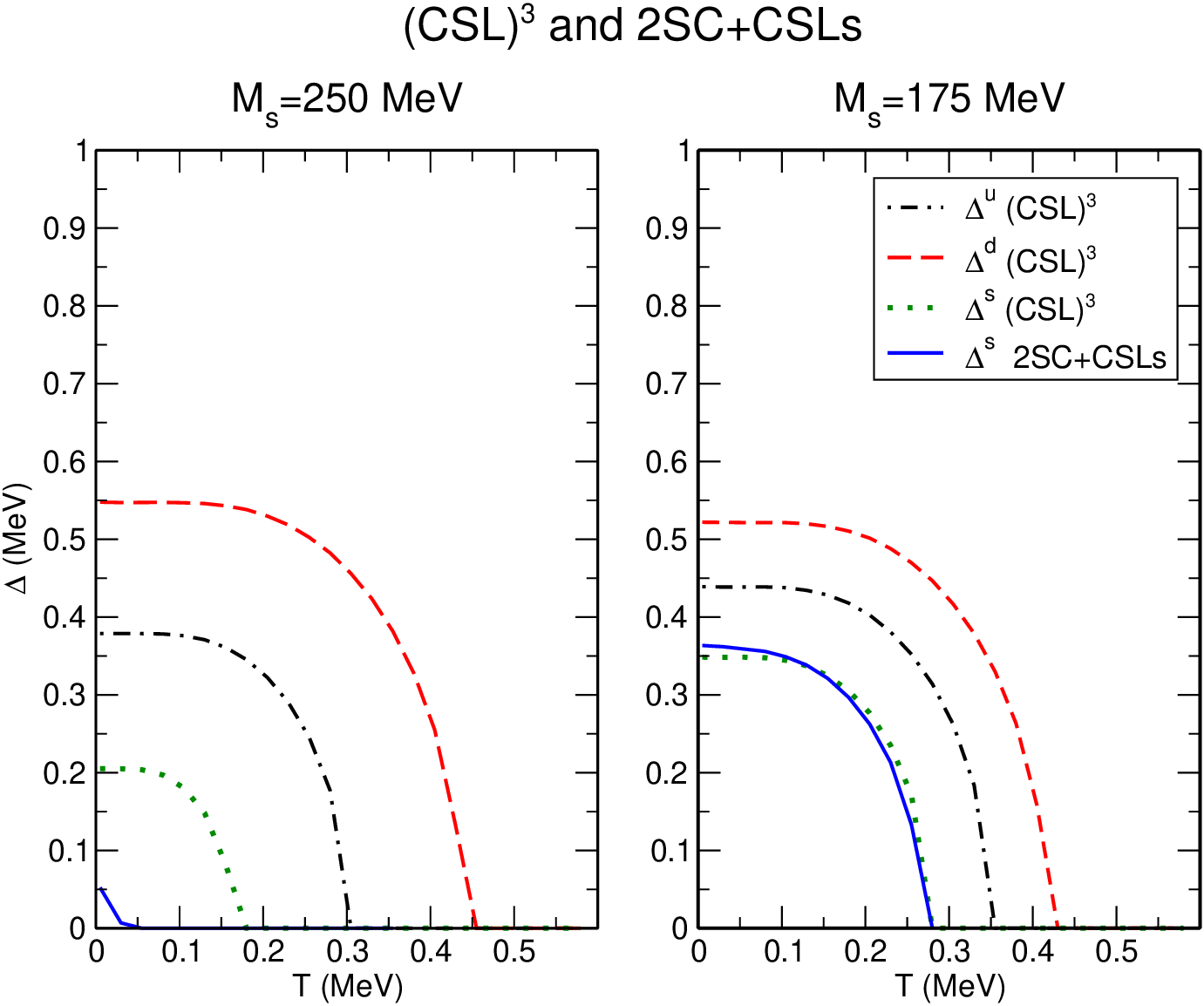}
  \caption{Variation of the 1SC (upper)
    and CSL (lower) gaps with temperature at $\mu=540$ MeV and for
    $M_s=250,175$ MeV. 
    Solid lines are for the strange quarks, in the case where up and down
quarks are undergoing 2SC pairing. The various broken lines are for
    the three flavours of quark in the case where all of them 
potentially undergo single-flavour pairing.}
  \label{fig:1SC_CSL_gaps_T}
\end{figure}

The temperature dependence of the single-flavour gap
parameters is shown in
Fig.~\ref{fig:1SC_CSL_gaps_T}. We see that the gap
parameters drop with
increasing temperature, and that each vanishes at a second-order
phase transition at some critical temperature $T_{\rm crit}$.
The critical temperatures deviate slightly from the BCS prediction
$T_{\rm crit} \simeq 0.57\, \De(T=0)$: 
\begin{eqnarray}
  \label{eq:CSL_T_dep}
\mbox{1SC}:\quad  T^f_{crit} &\approx& 0.51\, \De^f_{1SC}(T=0),\\
\mbox{CSL}:\quad  T^f_{crit} &\approx& 0.82\, \De^f_{\rm CSL}(T=0).
\end{eqnarray}
Such deviation was also found
in weak-coupling-QCD studies of 
single-flavour pairing \cite{Schmitt:2002sc}, and
non-BCS critical temperatures have also been seen
in gapless phases \cite{Huang:2003xd}
and crystalline phases \cite{Bowers:2001ip}.

\begin{figure*}[t]
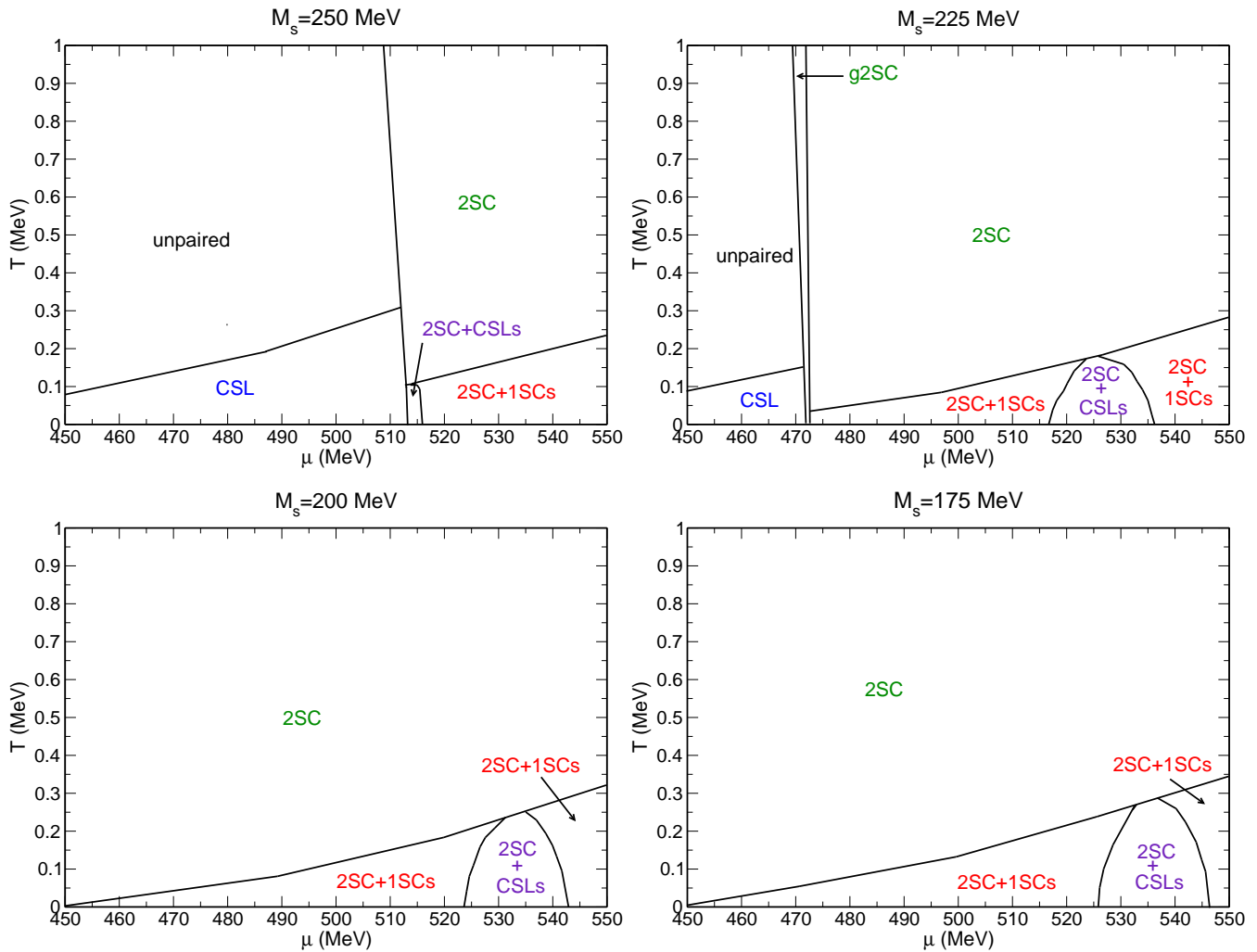

  \includegraphics[scale=0.36]{./figures/phase_diagram_Ms250.eps}
  \includegraphics[scale=0.36]{./figures/phase_diagram_Ms225.eps}\\[2ex]
  \includegraphics[scale=0.36]{./figures/phase_diagram_Ms200.eps}
  \includegraphics[scale=0.36]{./figures/phase_diagram_Ms175.eps}
  \caption{Phase diagrams for one- and two- flavour pairing in
  neutral three flavour quark matter with values of $M_s$ between
 175 and 250 MeV, calculated in an NJ model with cutoff $\La=600~\MeV$.
  For the \twoonesc\ and \twocsl\ phases see Figs.~\ref{fig:2sc_levels}
  and \ref{fig:2SC_CSL_levels}. The phase labelled ``CSL'' involves
  CSL pairing of some or all of the $u$,$d$, and $s$ quarks.
  }
  \label{fig:phase_diags}
\end{figure*}

It is interesting to note that because each flavour has a different
critical temperature for its self-pairing, we expect a series of
phase transitions as the system cools (imagining these phases 
to be in the core of a neutron star, for example), with single-flavour
pairing appearing first for the down quarks, then for the up quarks,
and finally for the strange quarks.

In the absence of any 2SC pairing, we also see that the critical temperature
for 1SC pairing of any given flavour is approximately equal to
the critical temperature for CSL pairing of the same flavour, 
even though the 1SC phase has a larger gap. This means that
if we heat up a \csl\ phase, we do not expect it to turn into
\onesc.
However, in the presence of 2SC pairing, we find that the critical
temperature for \twocsl\ is smaller than that for \twoonesc,
so we expect to find \twocsl$\to$\twoonesc$\to$2SC as we
heat up the system, and this is seen in the phase diagrams
of section \ref{sec:phase_diags}.

We repeated our calculations for a higher cutoff $\Lambda=800$~MeV,
calibrating the coupling to give the same 2SC gap, and we found that
the single-flavour pairing gaps dropped by a factor of $\sim 8$,
but the relations \eqn{eq:CSL_T_dep} were still valid.

\section{Discussion of phase diagrams}
\label{sec:phase_diags}

Section \ref{sec:model} describes how we calculate the
free energy for each phase as a function of ($\mu, T, M_s$). 
At each value of ($\mu$,$T$) the
favored phase is the one with the lowest free energy.
Fig. \ref{fig:phase_diags} presents this information as
a series of phase diagrams.

The crude structure of the diagrams involves
the expected transition from unpaired quark
matter at low density to two-flavour-paired quark matter (2SC) at
high density, at a critical chemical potential $\mu_{\rm crit} 
\approx M_s^2/(2\mu)$. In the $\mu$-$T$ diagram this transition 
occurs along a line that comes down very steeply at $\mu=\mu_{\rm crit}$,
because we are interested in low temperatures $T\lesssim 1~\MeV$,
so $T\ll\De_{2SC}$. As the strange quark mass drops, this line
moves to the left, and 2SC occupies more and more of the phase space.

In both the unpaired and 2SC regions of phase space, we find
that, at sufficiently low temperatures, single-flavour
pairing occurs. Below $\mu_{\rm crit}$
at zero temperature, the quark matter is in the \csl\ phase
in which all three flavours undergo CSL pairing, as depicted in
Fig.~\ref{fig:CSL_levels}.
Since the three flavours have different Fermi momenta,
they have different CSL gaps, and different critical temperatures,
so the region we have labelled in the phase diagrams
as ``CSL'' actually contains three
bands, one where just down quarks undergo CSL pairing, then below that
one where down and up quarks undergo CSL pairing, and finally
at the lowest temperatures one where all three flavours
undergo CSL pairing. We do not expect any 1SC pairing in this region
because the 1SC phases do not have higher critical temperatures
than the CSL phases (Fig.~\ref{fig:1SC_CSL_gaps_T}).

Above $\mu_{\rm crit}$, the phase structure is more complicated.
A colour chemical potential $\mu_8$ is generated in order to keep
the 2SC phase colour neutral, and this acts to 
split the blue quark Fermi momentum apart from the red and green quark
Fermi momenta. The value of $\mu_8$ is typically of order a few
MeV, which is easily large enough to disrupt any attempt at
CSL pairing, which itself only has a gap on the order of one MeV.
As a result, the favored phase at low temperature for
$\mu>\mu_{\rm crit}$ is typically the \twoonesc\ phase, in which the red and
green strange quarks form a spin-1 condensate (Fig.~\ref{fig:2sc_levels}).
At a particular value of $\mu$, however, $\mu_8$ may pass through
zero, and in that region of phase space, where $|\mu_8|< 2\De_{\rm CSL}$,
the strange quarks of all three colours are able to pair in a \twocsl\ phase
(Fig.~\ref{fig:2SC_CSL_levels}), which produces an ``island'' of
\twocsl\ in the \twoonesc\ band.
For our particular choice of NJL
model and cutoff, this occurs at $\mu\approx 530~\MeV$
for $M_s=175,\, 200,\, 225~\MeV$. As we emphasized in section 
Sect.~\ref{sec:chem_pots}, however, the position of such an
island is sensitive to details of the cutoff, and is not a
robust prediction of the model.

\section{Conclusions}
\label{sec:csl_summ}

In this paper we have shown that single flavour pairing
is strongly affected by what is happening to other flavours.
In particular, two-flavour pairing of the light flavours
induces large enough colour chemical potentials to change
the nature of the favored single-flavour pairing pattern
for the strange quarks.
Our results, shown in Fig.~\ref{fig:phase_diags},
were obtained using a NJL model
in the mean field approximation
with a pointlike four-fermion interaction whose
index structure is based on single gluon exchange.
We treated the up and down quarks as massless, and treated the
strange quark mass as a fixed parameter.
Some of the interesting features, such as the \twocsl\ ``island'',
are cutoff-dependent, but we found that
the basic structure of the diagram is a robust prediction
of the model. At high temperatures ($T\gg 1~\MeV$), only flavour-antisymmetric
pairing such as 2SC can survive. At
sufficiently low temperatures we find either CSL pairing of
some or all flavours,
or \twoonesc\ pairing. The critical temperature for the formation of these
condensates is not reliably predicted by our model, but appears to
be of order 1 MeV.

Our analysis was focused on the 2SC/1SC/CSL comparison.
We ignored other single-flavour pairing patterns
such as those with Dirac structure $C\si_{03}$ and $C\ga_0\ga_5$, since for 
an NJL model with our interaction
they have much smaller gap parameters, and hence either make a negligible
contribution to the free energy, or are
disfavored relative to the 1SC ($C\ga_3$) phase \cite{Alford:2002kj}.
(It is interesting to note that weak-coupling QCD calculations
find that a transverse {\em planar} phase is the strongest competitor
to CSL \cite{Schmitt:2004et}; this phase has not
yet been studied in the NJL context.)
More importantly, we also 
explicitly ignored other possible competing phases such as the
CFL phase and crystalline (LOFF) pairing, and we also
ignored chiral condensates
that would give larger (density dependent) quark masses.
One natural direction in which to extend this work is to perform
a simultaneous comparison of all known phases, along the lines of
Refs.~\cite{Ruster:2005jc,Blaschke:2005uj}, although a definitive calculation
is not yet possible because we still do not know what true ground
state underlies the unstable gapless phases 
\cite{Huang:2003xd,Alford:2003fq,Huang:2004am,Casalbuoni:2004tb,Giannakis:2004pf,Fukushima:2005cm}.

Another topic for future work is the possible ramifications of
our results for the phenomenology of quark matter in neutron stars.
The phase diagrams of Fig.~\ref{fig:phase_diags} show that a core
of quark matter inside a neutron star could have a complicated
radial structure of different phases, each with its own
special properties. Moreover, this structure would change over time
as the star's interior cooled. In our calculation the single-flavour
phases had gaps of order 1~MeV, so the phase structure would settle down
less than a year after the supernova,
but it should be noted that we completely ignored
the very weak single-flavour single-colour pairing that accompanies
most of these phases, whose gap parameter could be as small as 1 eV
\cite{Alford:2002kj}, leading to the emergence of further phase structure
after millions of years.

For neutron star phenomenology, we are interested in the transport properties 
of the different phases, in particular their permeability to magnetic
fields, as well as thermal properties, neutrino emissivity and 
mean free path, viscosity, and so on. These are sensitive to the
low-energy degrees of freedom, such as Goldstone bosons and
ungapped quark modes. 
We postpone a detailed study
and now sketch some of the obvious features.
First, some general points:
a massive flavour with CSL pairing is completely 
and isotropically gapped with an energy gap of approximately
$m\Delta/\mu$ for $m,\De\ll\mu$, so some of the quasiquarks become
gapless in the massless limit (see discussion after Eq.~\eqn{eq:CSLdet}).
A massive flavor with 1SC pairing has a direction-dependent gap, and in the
massless limit becomes ungapped at two points on the Fermi surface.
However we must remember that, like 2SC pairing,
1SC pairing only involves two of the three colours, so there may
be other gapless modes involving the third colour.
Focussing on the phases that we have discussed here,
the most completely gapped phase is \csl, in which all quark modes
are gapped, although if the up and down flavours have low
constituent masses there will be modes with rather small gaps.
(From Ref.~\cite{Ruster:2005jc}, Fig.~4, we expect $m_u,m_d\sim 20~\MeV$
at $\mu\approx 400~\MeV$, so assuming a CSL gap parameter of
1~\MeV\ we find an energy gap of about $50~\keV$.)
Next comes the \twocsl\ phase, which 
has blue up and down quarks that are effectively
gapless (they may have an eV-scale gap due to pairing in some 
very weak channel such as
$C\ga_0\ga_5$, which is suppressed for light quarks). 
Finally, the \twoonesc\ phase has in addition the blue strange
quarks (which can generate a bigger $C\ga_0\ga_5$ self-pairing gap
because the strange quark is heavier). 
The \twoonesc\ phase is also special because
rotational invariance is
broken by the 1SC pairing, and for the red and green strange quarks
the gap varies over the Fermi surface, but is always of order 
$\De_{\rm 1SC}^s$ \cite{Alford:2002kj}.
The full cooling phenomenology
of these phases has not yet been worked out, although
such questions have already been discussed
for isolated single flavours \cite{Schmitt:2005wg} and for
two-flavour quark matter \cite{Aguilera:2005tg}.

The magnetic properties of these phases are also expected to show
some variety.
In unpaired quark matter there is no Meissner effect,
so magnetic flux is not expelled or confined to flux tubes.
In an idealized 2SC phase the four participating quark species
are uniformly gapped, but the remainder are gapless.
The colour $SU(3)$ and electromagnetic $U(1)$ gauge symmetries are
broken down to a colour $SU(2)$ (giving masses to five of the gluons)
and a rotated electromagnetic $U(1)_{\tilde Q}$.
In the \csl\ phase all the gauge bosons have Meissner masses. 
The gluon masses are of order $g \De_{\rm CSL}$, and the photon mass
is of order $e \De_{\rm CSL}$ \cite{Schmitt:2003xq,Schmitt:2003aa}. The
magnetic properties of the \twocsl\ phase have not yet been
calculated: we expect mixing of the photon and the eighth gluon,
with Meissner masses that depend on $g$ as well as $e$.
In the \twoonesc\ phase, under the $U(1)_{\tilde Q}$
which survived the formation of the 2SC condensate,
the red, green, and blue strange quarks have charges
$-1,0,0$ respectively, so the additional 1SCs pairing
breaks the $U(1)_{\tilde Q}$ symmetry, giving the $\tilde Q$ photon
a mass of order $\tilde e\De_{\rm 1SC}^s$. The gluons have 
much larger masses of order $g \De_{\rm 2SC}$. It remains to be seen whether
this phase is a type-I or type-II superconductor for the
$\tilde Q$ photon.

We conclude that although single-flavour pairing makes a very small
contribution to the free energy of quark matter, it contributes
interesting extra structure to the low-temperature part of
the quark matter phase diagram. Single-flavour pairing determines
the symmetries and low-energy-excitations in this region, making
it directly relevant for the study of signatures of quark matter
in neutron stars.

\bc{\bf Acknowledgements}\ec
This research was supported by the U.K. Particle Physics
and Astronomy Research Council and by the U.S. Department of Energy
under contracts DE-FG02-91ER50628 and DE-FG01-04ER0225 (OJI).
We thank Andreas Schmitt for helpful comments and discussions.


\begin{thebibliography}{99}

\bibitem{BarroisPhD}
  B.~C.~Barrois,
  ``Nonperturbative Effects In Dense Quark Matter,'' Ph.D. thesis
  UMI 79-04847.

\bibitem{Bailin:1983bm}
  D.~Bailin and A.~Love,
  Phys.\ Rept.\  {\bf 107}, 325 (1984).

\bibitem{ARW2}
  M.~G.~Alford, K.~Rajagopal and F.~Wilczek,
  Phys.\ Lett.\ B {\bf 422}, 247 (1998)
  [arXiv:hep-ph/9711395].

\bibitem{Rapp:1997zu}
  R.~Rapp, T.~Schafer, E.~V.~Shuryak and M.~Velkovsky,
  Phys.\ Rev.\ Lett.\  {\bf 81}, 53 (1998)
  [arXiv:hep-ph/9711396].

\bibitem{Rajagopal:2000wf}
  K.~Rajagopal and F.~Wilczek,
  arXiv:hep-ph/0011333.

\bibitem{alford_review}
  M.~G.~Alford,
  Ann.\ Rev.\ Nucl.\ Part.\ Sci.\  {\bf 51}, 131 (2001)
  [arXiv:hep-ph/0102047].

\bibitem{Alford:2002ng}
M.~G.~Alford,
Nucl.\ Phys.\ Proc.\ Suppl.\  {\bf 117} (2003) 65
[arXiv:hep-ph/0209287].

\bibitem{Nardulli:2002ma}
G.~Nardulli,
Riv.\ Nuovo Cim.\  {\bf 25N3} (2002) 1
[arXiv:hep-ph/0202037].

\bibitem{Schafer:2003vz}
T.~Schafer,
arXiv:hep-ph/0304281.

\bibitem{Glendenning} 
  N. K. Glendenning, 
  ``Compact stars: Nuclear physics, particle physics, and general relativity'',
  Springer-Verlag,  New York, USA (1997).

\bibitem{Senger:2005pu}
  P.~Senger,
  J.\ Phys.\ G {\bf 31}, S1111 (2005).

\bibitem{bcs1}
J. Bardeen, L.N. Cooper, J.R. Schrieffer, Phys. Rev. \textbf{106}, 162 (1957)

\bibitem{Buballa:2002xv}
  M.~Buballa and M.~Oertel,
  arXiv:hep-ph/0205027.

\bibitem{ARW3}
  M.~G.~Alford, K.~Rajagopal and F.~Wilczek,
  Nucl.\ Phys.\ B {\bf 537}, 443 (1999)
  [arXiv:hep-ph/9804403].

\bibitem{Alford:1999pa}
M.~G.~Alford, J.~Berges and K.~Rajagopal,
Nucl.\ Phys.\ B {\bf 558} (1999) 219
[arXiv:hep-ph/9903502].

\bibitem{Buballa:2001gj}
M.~Buballa and M.~Oertel,
Nucl.\ Phys.\ A {\bf 703} (2002) 770
[arXiv:hep-ph/0109095].

\bibitem{IwaIwa}
  M.~Iwasaki and T.~Iwado,
  Phys.\ Lett.\ B {\bf 350}, 163 (1995).
  M.~Iwasaki and T.~Iwado,
  Prog.\ Theor.\ Phys.\  {\bf 94}, 1073 (1995).
  M.~Iwasaki, S.~Ishikawa and T.~Tanaka,
  Prog.\ Theor.\ Phys.\  {\bf 104}, 1041 (2000).

\bibitem{TS1flav}
  T.~Schafer,
  Phys.\ Rev.\ D {\bf 62}, 094007 (2000)
  [arXiv:hep-ph/0006034].

\bibitem{BHO}
M.~Buballa, J.~Hosek and M.~Oertel,
Phys.\ Rev.\ Lett.\  {\bf 90} (2003) 182002
[arXiv:hep-ph/0204275].

\bibitem{Alford:2002kj}
M.~G.~Alford, J.~A.~Bowers, J.~M.~Cheyne and G.~A.~Cowan,
Phys.\ Rev.\ D {\bf 67} (2003) 054018
[arXiv:hep-ph/0210106].

\bibitem{Schmitt:2004hg}
  A.~Schmitt,
  arXiv:nucl-th/0405076.

\bibitem{Schmitt:2004et}
A.~Schmitt,
Phys.\ Rev.\ D {\bf 71} (2005) 054016
[arXiv:nucl-th/0412033].

\bibitem{Alford:2003fq}
M.~Alford, C.~Kouvaris and K.~Rajagopal,
Phys.\ Rev.\ Lett.\  {\bf 92} (2004) 222001
[arXiv:hep-ph/0311286].

\bibitem{Alford:2004hz}
  M.~Alford, C.~Kouvaris and K.~Rajagopal,
  Phys.\ Rev.\ D {\bf 71}, 054009 (2005)
  [arXiv:hep-ph/0406137].

\bibitem{OurLOFF}
M.~Alford, J.~Bowers and K.~Rajagopal,
Phys.\ Rev.\ D {\bf 63}, 074016 (2001)
[hep-ph/0008208].

\bibitem{Bowers:2002xr}
J.~A.~Bowers and K.~Rajagopal,
Phys. Rev. D {\bf 66}, 065002 (2002) [hep-ph/0204079].

\bibitem{Bowers_thesis} 
J.~A.~Bowers,
arXiv:hep-ph/0305301.


\bibitem{Ruster:2005jc}
  S.~B.~Ruster, V.~Werth, M.~Buballa, I.~A.~Shovkovy and D.~H.~Rischke,
  Phys.\ Rev.\ D {\bf 72}, 034004 (2005)
  [arXiv:hep-ph/0503184].


\bibitem{Abuki:2005ms}
  H.~Abuki and T.~Kunihiro,
  arXiv:hep-ph/0509172.

\bibitem{Steiner:2002gx}
A.~W.~Steiner, S.~Reddy and M.~Prakash,
Phys.\ Rev.\ D {\bf 66} (2002) 094007
[arXiv:hep-ph/0205201].

\bibitem{Ruster:2005ib}
  S.~B.~Ruster, V.~Werth, M.~Buballa, I.~A.~Shovkovy and D.~H.~Rischke,
  arXiv:hep-ph/0509073.

\bibitem{Huang:2001yw}
M.~Huang, P.~f.~Zhuang and W.~q.~Chao,
Phys.\ Rev.\ D {\bf 65} (2002) 076012
[arXiv:hep-ph/0112124].

\bibitem{Blaschke:2003cv}
D.~Blaschke, M.~K.~Volkov and V.~L.~Yudichev,
arXiv:hep-ph/0301065.

\bibitem{helium3} 
D.~Vollhardt and P.~W\"{o}lfle, The Superfluid
  Phases of Helium 3, Taylor and Francis, (1990).

\bibitem{Gerhold:2003js}
A.~Gerhold and A.~Rebhan,
Phys.\ Rev.\ D {\bf 68} (2003) 011502
[arXiv:hep-ph/0305108].


\bibitem{Dietrich:2003nu}
D.~D.~Dietrich and D.~H.~Rischke,
Prog.\ Part.\ Nucl.\ Phys.\  {\bf 53} (2004) 305
[arXiv:nucl-th/0312044].

\bibitem{Gerhold:2004sk}
A.~Gerhold,
arXiv:hep-ph/0411086.

\bibitem{Buballa:2005bv}
  M.~Buballa and I.~A.~Shovkovy,
  Phys.\ Rev.\ D {\bf 72}, 097501 (2005)
  [arXiv:hep-ph/0508197].

\bibitem{Bedaque:2001je}
P.~F.~Bedaque and T.~Schafer,
Nucl.\ Phys.\ A {\bf 697}, 802 (2002)
[arXiv:hep-ph/0105150];

\bibitem{Neumann:2002jm}
F.~Neumann, M.~Buballa and M.~Oertel,
Nucl.\ Phys.\ A {\bf 714} (2003) 481
[arXiv:hep-ph/0210078].

\bibitem{Reddy:2004my}
  S.~Reddy and G.~Rupak,
  Phys.\ Rev.\ C {\bf 71}, 025201 (2005)
  [arXiv:nucl-th/0405054].

\bibitem{Kikkawa:1976fe}
  K.~Kikkawa,
  Prog.\ Theor.\ Phys.\  {\bf 56}, 947 (1976).

\bibitem{Ebert:1991pz}
D.~Ebert, L.~Kaschluhn and G.~Kastelewicz,
Phys.\ Lett.\ B {\bf 264} (1991) 420.

\bibitem{Reinhardt:1989rw}
H.~Reinhardt,
Phys.\ Lett.\ B {\bf 244} (1990) 316.

\bibitem{Berges:1998rc}
J.~Berges and K.~Rajagopal,
Nucl.\ Phys.\ B {\bf 538} (1999) 215
[arXiv:hep-ph/9804233].

\bibitem{Shovkovy:2003uu}
I.~Shovkovy and M.~Huang,
Phys.\ Lett.\ B {\bf 564} (2003) 205
[arXiv:hep-ph/0302142].

\bibitem{Huang:2003xd}
M.~Huang and I.~Shovkovy,
Nucl.\ Phys.\ A {\bf 729} (2003) 835
[arXiv:hep-ph/0307273].

\bibitem{Aguilera:2005tg}
  D.~N.~Aguilera, D.~Blaschke, M.~Buballa and V.~L.~Yudichev,
  Phys.\ Rev.\ D {\bf 72}, 034008 (2005)
  [arXiv:hep-ph/0503288].

\bibitem{Schmitt:2005wg}
  A.~Schmitt, I.~A.~Shovkovy and Q.~Wang,
  arXiv:hep-ph/0510347.

\bibitem{AR-02}
M.~Alford and K.~Rajagopal,
JHEP {\bf 0206} (2002) 031
[hep-ph/0204001].

\bibitem{Schmitt:2002sc}
A.~Schmitt, Q.~Wang and D.~H.~Rischke,
Phys.\ Rev.\ D {\bf 66} (2002) 114010
[arXiv:nucl-th/0209050].

\bibitem{Bowers:2001ip}
J.~A.~Bowers, J.~Kundu, K.~Rajagopal and E.~Shuster,
Phys.\ Rev.\ D {\bf 64} (2001) 014024
[arXiv:hep-ph/0101067].

\bibitem{Blaschke:2005uj}
  D.~Blaschke, S.~Fredriksson, H.~Grigorian, A.~M.~Oztas and F.~Sandin,
  Phys.\ Rev.\ D {\bf 72}, 065020 (2005)
  [arXiv:hep-ph/0503194].

\bibitem{Huang:2004am}
  M.~Huang and I.~A.~Shovkovy,
  Phys.\ Rev.\ D {\bf 70}, 094030 (2004)
  [hep-ph/0408268].

\bibitem{Casalbuoni:2004tb}
  R.~Casalbuoni, R.~Gatto, M.~Mannarelli, G.~Nardulli and M.~Ruggieri,
  Phys.\ Lett.\ B {\bf 605}, 362 (2005)
  [Erratum-ibid.\ B {\bf 615}, 297 (2005)]
  [hep-ph/0410401].

\bibitem{Giannakis:2004pf}
  I.~Giannakis and H.~C.~Ren,
  Phys.\ Lett.\ B {\bf 611}, 137 (2005)
  [hep-ph/0412015].

\bibitem{Fukushima:2005cm}
  K.~Fukushima,
  hep-ph/0506080.

\bibitem{Schmitt:2003xq}
  A.~Schmitt, Q.~Wang and D.~H.~Rischke,
  Phys.\ Rev.\ Lett.\  {\bf 91} (2003) 242301
  [arXiv:nucl-th/0301090].

\bibitem{Schmitt:2003aa}
  A.~Schmitt, Q.~Wang and D.~H.~Rischke,
  Phys.\ Rev.\ D {\bf 69} (2004) 094017
  [arXiv:nucl-th/0311006].

\end{thebibliography}
\end{document}